\documentclass[journal]{IEEEtran}

\usepackage{subcaption}
\usepackage{tabularx}
\usepackage{epsfig}
\usepackage{hhline}
\usepackage{multirow}
\newcolumntype{C}[1]{>{\centering\let\newline\\\arraybackslash\hspace{0pt}}m{#1}}
\newcolumntype{K}[1]{>{\centering\arraybackslash}p{#1}}
\usepackage{amsmath}

\begin{document}
\title{Separation of Instrument Sounds using Non-negative Matrix Factorization with Spectral Envelope Constraints}

\author{Jeongsoo Park, Jaeyoung Shin and Kyogu Lee,~\IEEEmembership{Senior Member,~IEEE}}

\maketitle
\begin{abstract}
Spectral envelope is one of the most important features that characterize the timbre of an instrument sound. However, it is difficult to use spectral information in the framework of conventional spectrogram decomposition methods. We overcome this problem by suggesting a simple way to provide a constraint on the spectral envelope calculated by linear prediction. In the first part of this study, we use a pre-trained spectral envelope of known instruments as the constraint. Then we apply the same idea to a blind scenario in which the instruments are unknown. The experimental results reveal that the proposed method outperforms the conventional methods.
\end{abstract}


\section{Introduction}
\IEEEPARstart{S}{ource} separation in a digital signal processing aims to recover original signals of interest from given signal mixtures. It has been attracted considerable attention as a research topic in the past few decades and applied to many research fields \cite{adali2014source}. The applications of source separation include music and audio analysis such as instrument-wise equalizing, stereo-to-surround up-mixing, karaoke systems, and crosstalk cancellation, biomedical signal analysis such as electroencephalographic (EEG) and electromyographic (EMG) \cite{delorme2004eeglab, hesse2006semi, vazquez2012blind}, and chemical signal analysis \cite{duarte2014source}. Nevertheless, the mainstream of the recent source separation research focuses on the audio signal due to the easily overlapping nature of sound and its diverse applications.

In almost every situation we hear a variety of sounds that occur simultaneously, and humans are able to find meaningful information in the sounds. ``Cocktail party problem'' demonstrates an interesting phenomenon concerning the human's capability of listening. Even in a noisy environment like the cocktail party, people are able to concentrate on a sound that they want to attend such as the voice of a person they are in conversation. This selective attention enables humans to catch crucial auditory information, with being insensitive to the magnitude of the sound. As this process happens in a human brain unconsciously, machines are not capable of imitating their magnitude-robust operation. Hence, in order to make the machines work correctly, pre-processing to amplify, or separate, the sounds of interest is necessary. This leads to the necessity of audio source separation, and this is why the source separation algorithms have an enormous impact on the audio signal processing and machine learning research. 

In particular, the source separation has been extensively used in the field of music signal analysis. As a pre-processing method, music source separation has contributed to the improvement of the various music information retrieval (MIR) algorithms enabling the extraction of musical information to work with high accuracy. Such algorithms include low-level music analyses such as pitch detection, tempo estimation, and instrument identification, and high-level analyses such as genre classification, music identification, and copyright monitoring \cite{ewert2014score}. Besides, it has also assisted music generation algorithms such as automatic music composition as well as music enhancement algorithms such as up-mixing, time-stretching, instrument-wise equalization, and noise reduction \cite{bryan2014interactive, driedger2016review}. In addition, the spectral and temporal characteristics of music signals are often considered stationary. Hence the algorithms to analyze the music signals can be extended to other applications to investigate noisier signals such as EEG signal.

\subsection{Number of channels}
The source separation research can be categorized according to the number channels of the input sound. When the mixture signal is composed of multiple channels, the estimation of original sources is achievable via spatial filtering. Conventional studies that use spatial filtering assume that the mixture signal is the linear combination \cite{li2006underdetermined} or convolutional mixing \cite{pedersen2008convolutive, ozerov2010multichannel} of the original sounds. In this case, the separation process is identical to obtaining the inverse of the actual mixing matrix. When the number of channels is equal to the number of original sources, the task is categorized as {\it determined} case since the perfect reconstruction of the original sounds is theoretically available. This is also applied to the {\it overdetermined} case where the number of channels exceeds the number of sources. However, when the number of channels is smaller than the number of sources, which is referred to as {\it underdetermined} case, the perfect reconstruction is not possible via spatial filtering; hence, the assistance of spectro-temporal characteristics is necessary.

Since most of the music signals are comprised of a single channel ({\it mono}) or two channels ({\it stereo}), the music source separation task is often presumed as underdetermined. Accordingly, intensive study about the single channel-based music source separation methods is essential, and it is commonly utilized as an essential background of the multi-channel music source separation \cite{aichner2007multi}.

\subsection{Utilization of side-information}
Meanwhile, the amount of information we use is considered as an important criterion to categorize source separation studies. In the early stage of the music source separation, the blind approach was intensively investigated \cite{virtanen2007monaural}, where no additional information about the target source exists. These blind source separation (BSS) studies often assume that the target sound has certain statistical features such as non-Gaussianity and independence \cite{davies2007source} or sparsity \cite{virtanen2007monaural}. BSS techniques are useful in some cases; however, such statistical assumptions cannot be guaranteed in many practical situations, which eventually causes performance degradation.

To overcome this low performance, informed source separation (ISS) was widely studied \cite{vincent2014blind}. These studies assume situations where side-information about the target sources is available. Such information includes spectro-temporal characteristics such as music score \cite{ewert2014score, ewert2012using, fritsch2013score}, partial information such as onset \cite{park2014separation}, and direct information such as manually provided annotations \cite{durrieu2012musical, lefevre2012semi, bryan2014isse} and user-guided audio signal \cite{smaragdis2009separation}. 

Especially, recent approaches have examined the effect of artificial neural network-based methods on the source separation tasks. While some of the research efforts directly applied the deep learning-based techniques like autoencoder as Lim and Lee's work \cite{limharmonic}, Osako {\it et al.}'s work \cite{osako2017supervised}, and Grais and Plumbley's work \cite{grais2017single}, others attempted to enhance conventional approaches like matrix decomposition \cite{kang2015nmf} and time-frequency mask \cite{huang2015joint}.

However, it is still a big issue to reduce the amount of information needed for successful separation of the target sound, as the situation where side-information can be provided is highly limited. Especially in deep learning-based approaches, a significant amount of training data is required for each sound source. Hence, we focus on reducing the amount of required information.

\subsection{Instrument sound separation}
Musical instruments are known to show static spectro-temporal characteristics and timbre. These characteristics can also be used for the source separation. Especially in cases where harmonic instruments and drums coexist, their attributes that appear in the time-frequency representation are easily distinguishable. Our previous work focused on the spectral features of harmonic and percussive sounds \cite{park2015harmonic}, and it was also extended to simultaneously consider the time and frequency domain aspects of the instruments \cite{park2017exploiting}. 

The separation of harmonic instrument sounds is a more challenging problem as the differences in the spectral and temporal characteristics are less obvious compared to the harmonic-percussive source separation. Spiertz and Gnann presented the basis clustering algorithm as a post-processing of non-negative matrix factorization (NMF) \cite{spiertz2009source}. Fitzgerald {\it et al.} used shift-invariant non-negative tensor factorization \cite{fitzgerald2008extended}. Ozerov and Fevotte focused on the mixing procedure \cite{ozerov2010multichannel}. 

Other studies on harmonic instrument sound separation adopt {\it source-filter model}. Heittola {\it et al.} trained bases for instruments and used them for separation \cite{heittola2009musical}. Rodriguez-Serrano {\it et al.} also made instrument-dependent models and used them for separation \cite{rodriguez2012multiple}. Klapuri {\it et al.}'s work extended Heittola {\it et al.}'s method to separately estimate approximated spectral envelopes and their corresponding excitations without the pre-training process \cite{klapuri2010sound}. But their work fails to precisely assess spectral envelopes since it roughly approximates the envelopes using band-pass filter banks. Ozerov {\it et al.} developed flexible audio source separation toolbox (FASST), in which spectra are also split into excitations and filter parts \cite{ozerov2012general, salaun2014flexible}.

In this paper, we present a novel approach to the single-channel instrument sound separation problem. In the first part of this paper, we assume that the instruments are known in advance and that their spectral envelopes can be obtained. For instance, when audio segments of the instruments are given, we can extract the spectral envelopes from them. The spectral bases of the NMF to approximate the mixture spectrogram are partitioned ahead of the iteration and then constrained to resemble the extracted envelopes. The constraint imposition is achieved with the use of generalized Dirichlet prior which has been used in the source separation tasks. In the second part of this paper, we extend the informed approach to the case where no additional information is available. In the blind approach, the spectral envelopes of each basis are calculated through LPC, and all envelopes belong to the same group are averaged. This is because the spectral envelope is determined for each instrument whereas the excitations can differ from other bases that belong to the same group. As the iteration proceeds, the average spectral envelopes of each group converge to the true spectral envelopes of the instruments. These two methods are based on the source-filter model with linear predictive coding (LPC), but they obtain the envelopes of the spectra (or spectral bases of the NMF) without converting them to the time-domain signal. The comparative evaluation reveals that the proposed method outperforms the other conventional methods. 

The rest of the paper is organized as follows. Section 2 describes our informed approach method. In Section 3, we extend the informed approach to blind approach. Section 4 shows the experimental results with real recordings. Conclusions are presented in Section 5.

\section{Proposed informed approach}
In this section, we present a detailed description of the proposed informed source separation method. This section is composed of three subsections that present excitation-filter model, linear predictive coding, and the proposed NMF-based spectrogram decomposition procedure. 

\subsection{Excitation-filter model}
The time domain mixture signal is generated by summing individual sources as
\begin{equation} 
\label{Eq1}
x = \sum\limits_{i = 1}^I {{x_i}}
\end{equation}
where $x$ denotes the mixture sound, $x_i$ denotes the sound of the $i$-th instrument, and $I$ denotes the number of instruments. When this time domain signal is converted into a spectrogram, it can be similarly represented as
\begin{equation} 
\label{Eq2}
\begin{aligned}
{\bf{X}} &= \sum\limits_{i = 1}^I {{{\bf{X}}_i}} \\
&\approx \sum\limits_{i = 1}^I {\sum\limits_{k \in {\Phi _i}} {{{\bf{w}}_k}{{\bf{h}}_k}} }
\end{aligned}
\end{equation}
where ${\bf{X}}_i$ denotes the magnitude spectrogram of instrument $i$, $\Phi _i$ denotes the index set of bases that explain ${\bf X} _i$, and ${{{\bf{w}}_k}}$ and ${{{\bf{h}}_k}}$ denote the $k$-th spectral basis and its time activation, respectively. The spectrogram conversion error is assumed to be small and negligible. For the convenience of description, we assume that ${|| {{{\bf{w}}_k}} ||_1} = 1$ for all $k$. ${\bf w}$ and ${\bf h}$ can be estimated with the matrix decomposition algorithms such as probabilistic latent semantic analysis (PLSA), probabilistic latent component analysis (PLCA) \cite{shashanka2008probabilistic}, and NMF \cite{lee1999learning}.

\begin{figure} 
\centering
\begin{subfigure}[]{0.48\textwidth}
\centering
\includegraphics[width=2.1in]{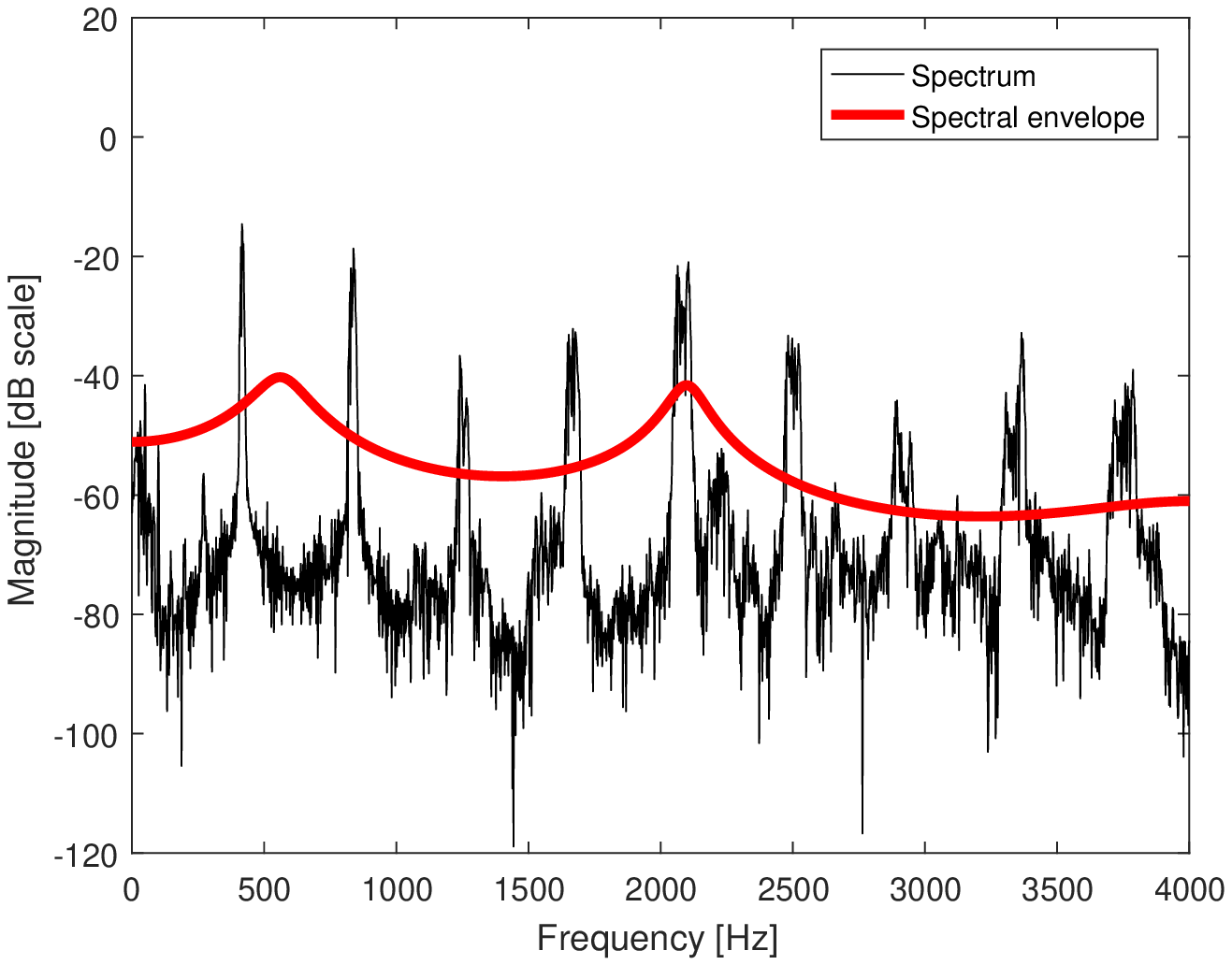}
\subcaption{}
\end{subfigure}
\begin{subfigure}[]{0.48\textwidth}
\centering
\includegraphics[width=2.1in]{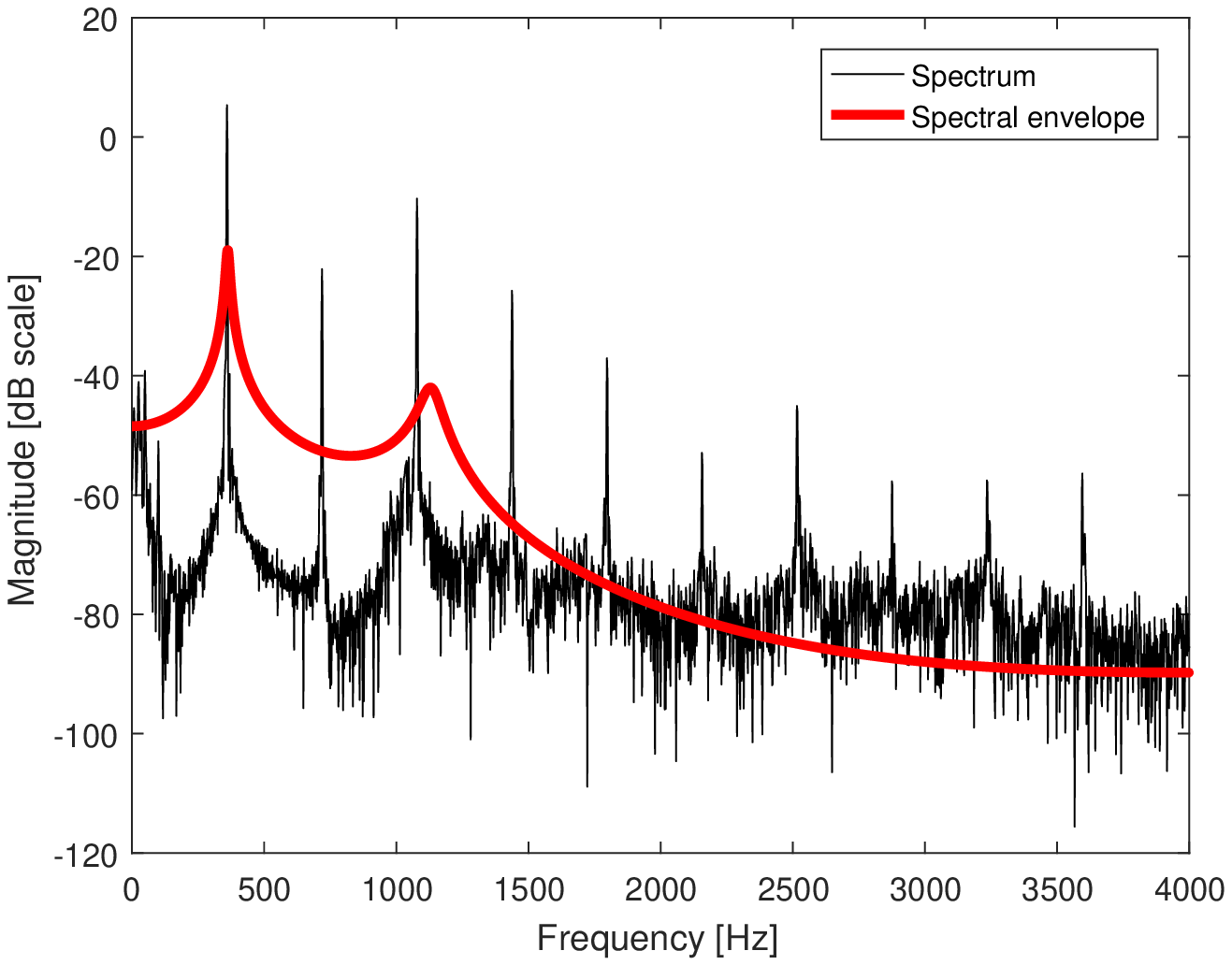}
\subcaption{}
\end{subfigure}

\begin{subfigure}[]{0.32\textwidth}
\centering
\includegraphics[width=2.1in]{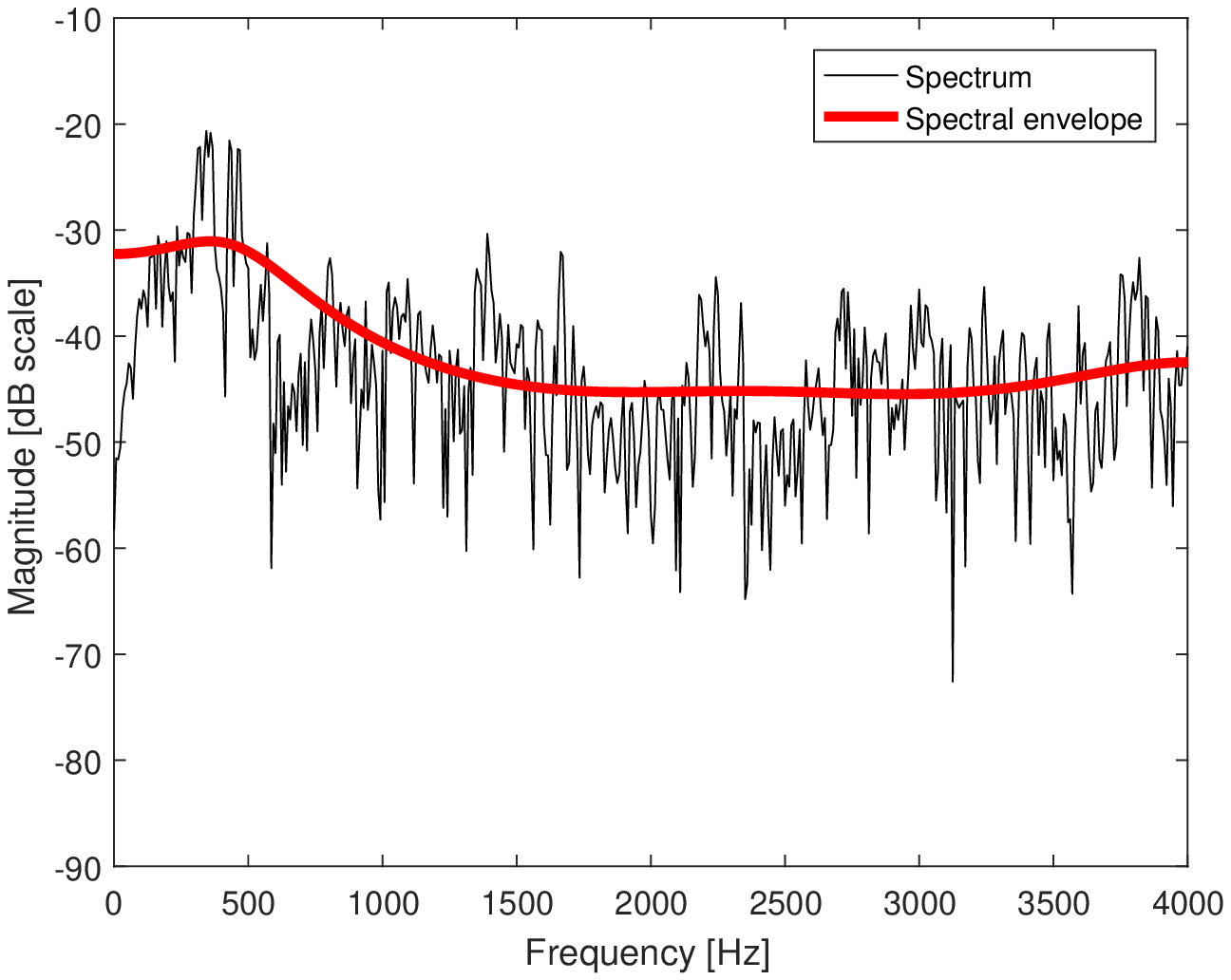}
\subcaption{}
\end{subfigure}
\begin{subfigure}[]{0.32\textwidth}
\centering
\includegraphics[width=2.1in]{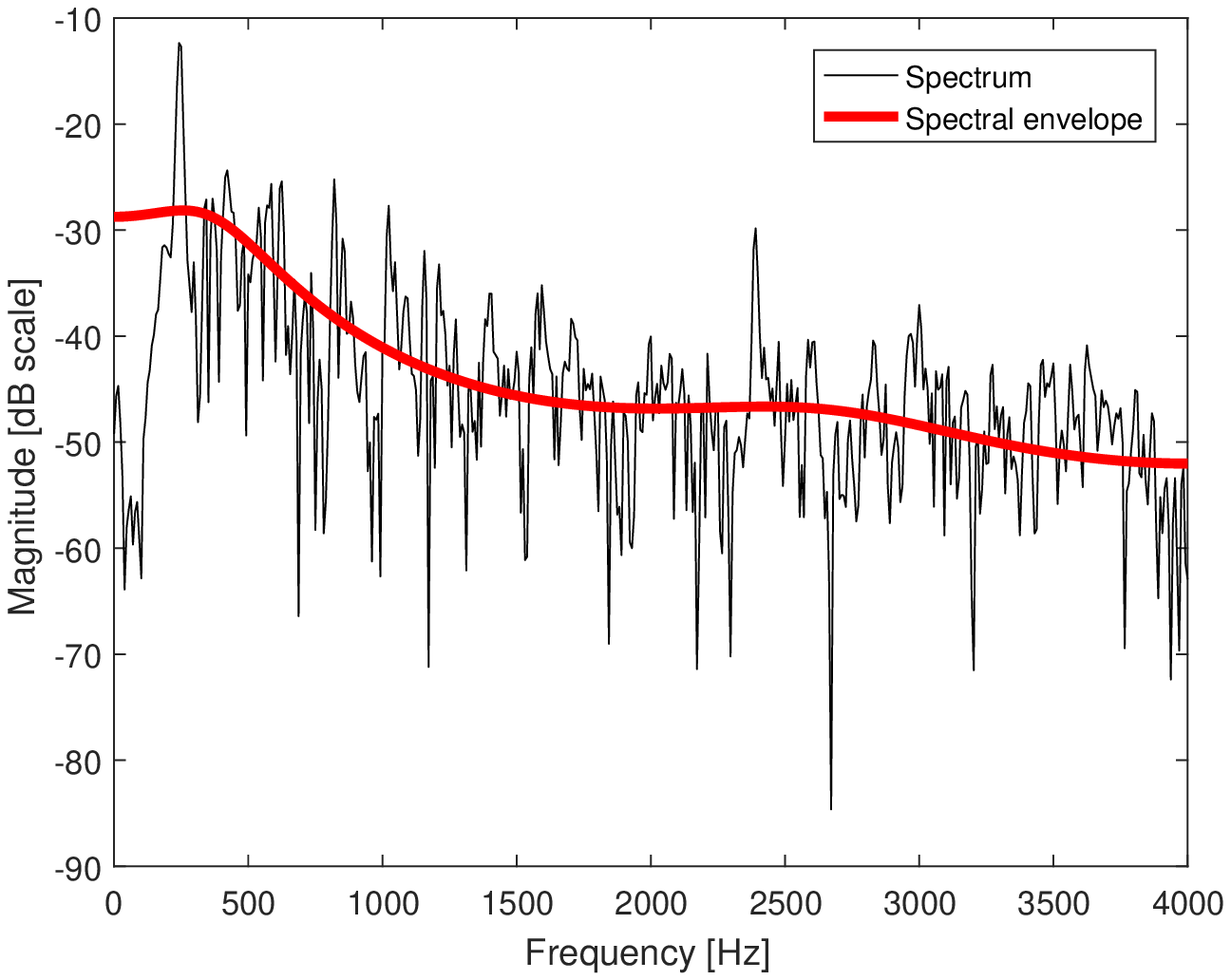}
\subcaption{}
\end{subfigure}
\begin{subfigure}[]{0.32\textwidth}
\centering
\includegraphics[width=2.1in]{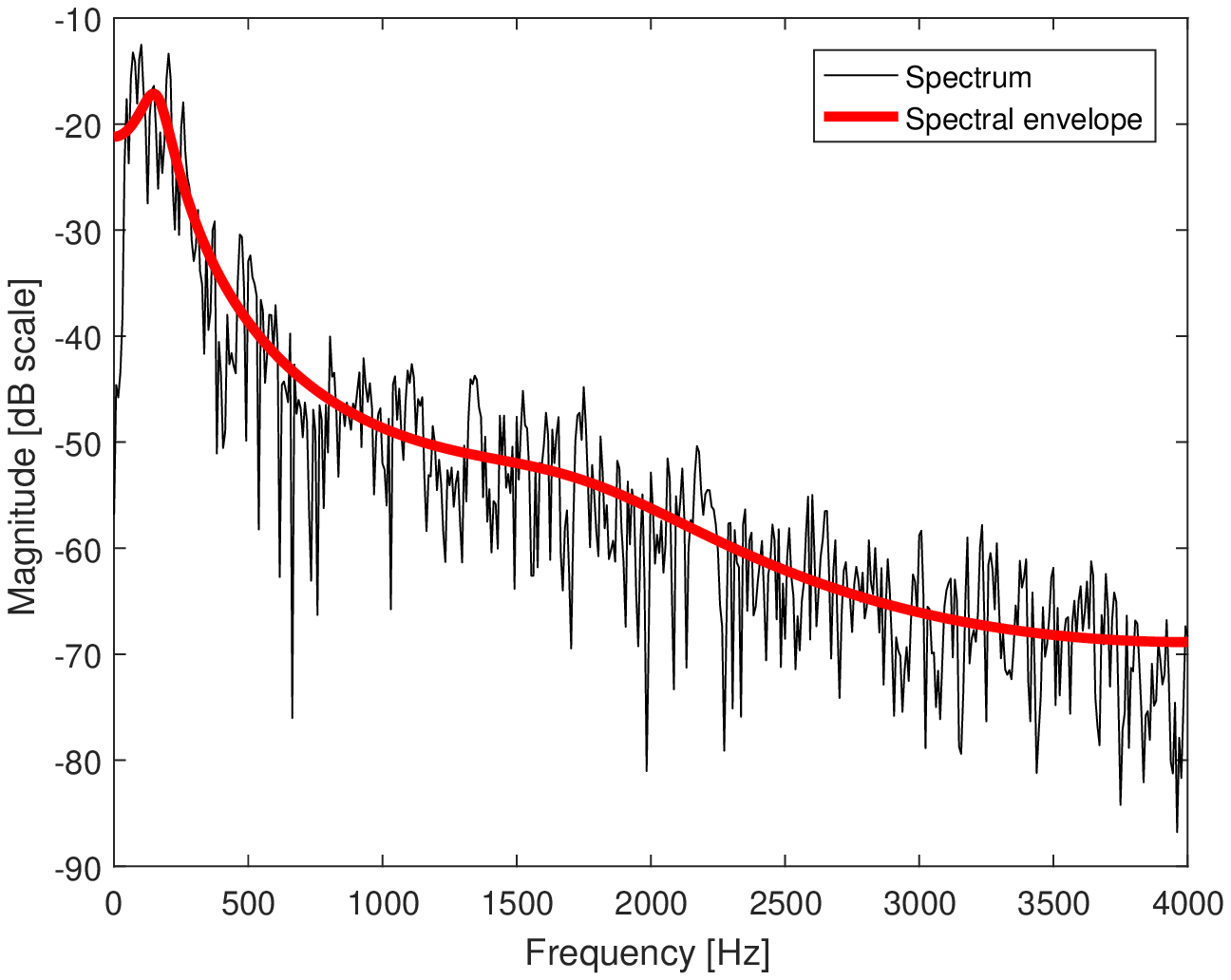}
\subcaption{}
\end{subfigure}
\caption{Spectrum and corresponding spectral envelope computed via linear prediction of (a) violin, (b) clarinet (c) hi-hat drum, (d) snare drum, and (e) bass drum.}
\label{Fig1}
\end{figure}

We also assume that the instrument sounds can be represented using the source-filter model, which we alternatively address as {\it excitation-filter model} to avoid term collision. The excitation-filter model has been widely used to analyze the speech production mechanism \cite{quatieri2006discrete}. According to the excitation-filter model, the timbre of an instrument is determined by its {\it filter}, whereas the pitch is determined by the excitation signal. Figure \ref{Fig1} (a) and (b) illustrate the spectrum and the corresponding LPC spectral envelope of violin and clarinet, respectively, plotted on a log-scale. Even though the excitation-filter model is to describe the generation of the harmonic instrument sound, we similarly extend it to the percussive instruments since they also present difference in the energy distribution of the spectra as shown in Figure \ref{Fig1} (c), (d), and (e). 

As the excitation signal is filtered by the instrument's resonant structure, a spectral basis (or a spectrum of an instrument, equivalently) ${{\bf w}_k}$ has to be represented as
\begin{equation}	
\label{Eq3}
{{\bf{w}}_k} = {{\bf{v}}_i} \odot {{\bf{e}}_k}
\end{equation}
where ${{\bf v}_i}$ is the filter's frequency response of instrument $i$ that can be alternatively referred to as {\it spectral envelope}, ${{\bf e}_k}$ is the spectrum of the $k$-th excitation signal, the operator $\odot$ denotes the element-wise multiplication, and $k \in {\Phi _i}$. 

Note that the above Eq. \ref{Eq3} can be satisfied only if a proper source separation is applied. During the ongoing matrix decomposition iteration, the spectral envelopes vary for each $k$. However, the proposed method focuses on the reverse direction of this theory: {\it Can a group of bases successfully reconstruct the sound of an instrument, if we can make the basis envelopes equal to the true envelope of the instrument?} Since the proposed method requires the estimation of the true spectral envelopes of the instruments and the basis envelopes, the representative spectral envelope extraction method is presented in the next subsection.

\subsection{Linear predictive coding}
The spectral envelope can be obtained using LPC, which assumes that the filter can be approximated by a finite number of poles. In this subsection, we describe how we can obtain LPC coefficients and how they can be extensively applied to the spectral bases of the NMF algorithm.

\subsubsection{Calculation of LPC coefficients}
LPC aims to calculate the all-pole infinite impulse response (IIR) filter coefficients ${\bf a} = {\left( {{a_1},{a_2},\,...\,,{a_M}} \right)^T}$ that best predict the signal value minimizing the energy of the error signal, represented as
\begin{equation} 
\label{Eq4}
\begin{aligned}
{\bf a} &= \arg \mathop {\min }\limits_{\bf{a}} E\left\{ {{{\left\Vert {err\left( n \right)} \right\Vert}^2}} \right\} \\ &= \arg \mathop {\min }\limits_{\bf{a}} E\left\{ {{{\left\Vert {y\left( n \right) - \sum\limits_{m = 1}^M {{a_m}y\left( {n - m} \right)} } \right\Vert^2}}} \right\},
\end{aligned}
\end{equation}
where $y$ is a real, time domain signal, $err$ is the error signal, and $M$ is the number of filter coefficients. The accurate estimation of LPC coefficients is important, since the filter's frequency response, namely spectral envelope, is represented as
\begin{equation} 
\label{Eq5}
H\left( z \right) = \frac{1}{{1 - \sum\limits_{m = 1}^M {{a_m}{z^{ - m}}} }}.
\end{equation}

The problem of computing LPC coefficients can be converted into an alternative form, which is referred to as {\it autocorrelation method}. It can be mathematically represented as
\begin{equation} 
\label{Eq6}
{{\bf R}}{\bf a} = {{\bf r}}
\end{equation}
where ${\bf r}$ is a vector of autocorrelations of $y$ defined as ${\bf r} = {\left( {{r_{yy}}\left( 1 \right),\,{r_{yy}}\left( 2 \right)\,,\,\,...\,\,,\,{r_{yy}}\left( M \right)} \right)^T}$, and ${\bf R}$ is an autocorrelation matrix defined as
\begin{equation} 
\label{Eq7}
{\bf{R}} = \left( {\begin{array}{*{20}{c}}
{{r_{yy}}\left( 0 \right)}&{{r_{yy}}\left( 1 \right)}& \cdots &{{r_{yy}}\left( {M - 1} \right)}\\
{{r_{yy}}\left( 1 \right)}&{{r_{yy}}\left( 0 \right)}&{}&{{r_{yy}}\left( {M - 2} \right)}\\
 \vdots &{}& \ddots & \vdots \\
{{r_{yy}}\left( {M - 1} \right)}&{{r_{yy}}\left( {M - 2} \right)}& \cdots &{{r_{yy}}\left( 0 \right)}
\end{array}} \right)
\end{equation}
where ${r_{yy}}\left( m \right) = E\left\{ {y\left( n \right) y \left( {n - m} \right)} \right\}$. From the above formula, we can observe that the calculation of LPC coefficients does not necessarily require the original time domain signal $y$ and that it can also be calculated by the autocorrelations. As ${\bf R}$ is a Toeplitz matrix, ${\bf a}$ can be easily obtained using Levinson-Durbin recursion \cite{vaseghi2008advanced}.

\subsubsection{Envelope of spectral bases}
Consider the spectrum ${\bf Y}$ of a time domain signal ${y\left( n \right)}$, and its magnitude ${\left| {\bf{Y}} \right|}$. The spectral envelope of the magnitude spectrum can be directly obtained without the spectrum-to-time domain signal conversion process. According to the Wiener-Khinchin theorem, the computation of autocorrelation is simplified as
\begin{equation} 
\label{Eq8}
\begin{aligned}
{{{\rho}}} &= IFFT\left[ {{{\bf{S}}_{yy}}} \right] \\
 &= IFFT\left[ {{\bf{Y}}{{\bf{Y}}^*}} \right] \\
 &= IFFT\left[ {{{\left| {\bf{Y}} \right|}^2}} \right]
\end{aligned}
\end{equation}
where ${{{\bf{S}}_{yy}}}$ is the power spectral density of ${y\left( n \right)}$, ${{\bf{r}}_{yy}}$ is defined as ${{{\rho}}} = {\left( {{r_{yy}}\left( 0 \right),\,{r_{yy}}\left( 1 \right),\,...\,,\,{r_{yy}}\left( M \right)} \right)^T}$, $IFFT\left[  \bullet  \right]$ denotes the inverse fast Fourier transform, and ${\left(  \bullet  \right)^*}$ denotes the complex conjugate. Here, we can see that the magnitude spectrum ${\left| {\bf{Y}} \right|}$ has sufficient information to attain the autocorrelations, which in turn can be used to estimate LPC coefficients. These LPC coefficients are finally used to estimate the spectral envelope.

\subsection{Spectrogram decomposition procedure}
On the basis of the equivalence of PLCA and NMF \cite{ding2008equivalence}, the application of Dirichlet to the NMF framework has been presented in some conventional works \cite{park2014separation, park2015harmonic, park2017exploiting}. According to those research, the Dirichlet prior is a way to shape bases in the probabilistic framework and can be easily generalized to the NMF framework. In this paper, we do not present how we can implement it on the PLCA framework. Instead, we directly describe our method on the NMF framework.

\begin{figure}[t]	
  \centering
  \centerline{\includegraphics[width=3.25in]{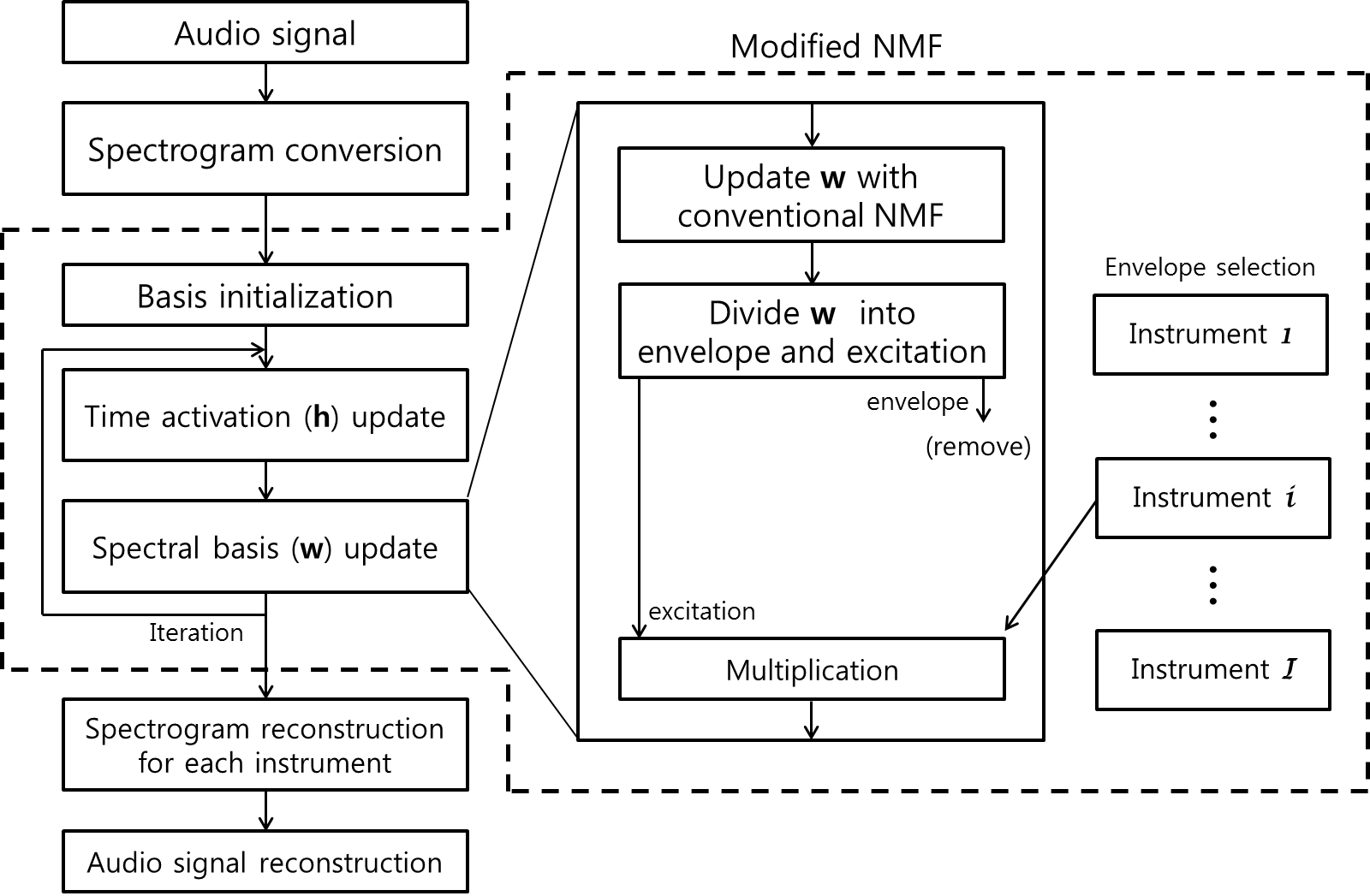}}
  \caption{Overview of the proposed informed approach.}
  \label{Fig2_overview}
\end{figure}

Figure \ref{Fig2_overview} illustrates the overall procedure of how the proposed method works. The input mixture audio is transformed into a magnitude spectrogram and is decomposed using the proposed modified NMF. In so doing, the bases are randomly initialized first, and the spectral bases and their corresponding time activations are estimated iteratively afterwards. After the estimation, ${\bf w}$ is divided into two parts, envelope ${\bf v}$ and excitation ${\bf e}$ using linear prediction. The envelopes of the bases that belong to an instrument's index set are replaced by the true envelope of the instrument. The spectral bases are then reconstructed by multiplying the new envelope and the excitation followed by the next iteration. After the iteration is finished, the spectrograms are reconstructed for each instrument. Finally, the audio signals are reconstructed.

The proposed method modifies the NMF update equations that minimize the KL divergence. It can be mathematically represented as
\begin{equation}	
\label{Eq9}
{\hat {\bf{H}}_{k,t}} \leftarrow \frac{{{\bf{H}}_{k,t}^{\left( l \right)}\sum\limits_f {\left\{ {{\bf{W}}_{f,k}^{\left( l \right)}{{{{\bf{X}}_{f,t}}} \mathord{\left/
 {\vphantom {{{{\bf{X}}_{f,t}}} {{{\tilde {\bf{X}}}_{f,t}}}}} \right.
 \kern-\nulldelimiterspace} {{{\tilde {\bf{X}}}_{f,t}}}}} \right\}} }}{{\sum\limits_{f'} {{\bf{W}}_{f',k}^{\left( l \right)}} }}
\end{equation}
\begin{equation}	
\label{Eq10}
{\tilde {\bf{W}}_{f,k}} \leftarrow \frac{{{\bf{W}}_{f,k}^{\left( l \right)}\sum\limits_t {\left\{ {{\bf{H}}_{k,t}^{\left( {l + 1} \right)}{{{{\bf{X}}_{f,t}}} \mathord{\left/
 {\vphantom {{{{\bf{X}}_{f,t}}} {{{\tilde {\bf{X}}}_{f,t}}}}} \right.
 \kern-\nulldelimiterspace} {{{\tilde {\bf{X}}}_{f,t}}}}} \right\}} }}{{\sum\limits_{t'} {{\bf{H}}_{k,t'}^{\left( {l + 1} \right)}} }}
\end{equation}
\begin{equation}	
\label{Eq11}
{\bf{H}}_{k,t}^{\left( {l + 1} \right)} \leftarrow {\hat {\bf{H}}_{k,t}}\sum\limits_f {{{\tilde {\bf{W}}}_{f,k}}}
\end{equation}
\begin{equation}	
\label{Eq12}
{\hat {\bf{W}}_{f,k}} \leftarrow \frac{{{{\tilde {\bf{W}}}_{f,k}}}}{{\sum\limits_f {{{\tilde {\bf{W}}}_{f,k}}} }}
\end{equation}
\begin{equation}	
\label{Eq13}
{\bf{w}}_k^{\left( {l + 1} \right)} \leftarrow \alpha {{\hat {\bf w}}_k} + \left( 1-\alpha \right) {{{\bf{\mathord{\buildrel{\lower3pt\hbox{$\scriptscriptstyle\smile$}} 
\over v} }}}^{\left( i \right)}} \odot {{\bf{e}}^{\left( k \right)}}, k \in \Phi_{i}
\end{equation}
where ${\bf H}$ is the $K \times N$ matrix of time activations, of which $k$-th row is ${{\bf h}_k}$, ${\bf W}$ is the $M \times K$ matrix of spectral bases, of which $k$-th column is ${{\bf w}_k}$, ${{\hat {\bf W}}}$ and $\tilde {\bf{W}}$ denote the temporarily adopted variables, which have the same size as ${\bf{W}}$, ${{\hat {\bf H}}}$ denotes the temporarily adopted variable that has the same size as $\bf{H}$, ${\tilde {\bf{X}}}$ is the estimated spectrogram reconstructed with up-to-date ${\bf W}$ and ${\bf H}$, ${{{\bf{v}}^{\left( k \right)}}} = {\left( {v_1^{\left( k \right)},\,...\,,\,v_F^{\left( k \right)}} \right)^T}$ and ${{\bf{e}}^{\left( k \right)}} = {\left( {e_1^{\left( k \right)},\,...\,,\,e_F^{\left( k \right)}} \right)^T}$ denote the spectral envelope and excitation spectrum of the $k$-th column of ${{\hat {\bf W}}}$ represented as ${{\hat {\bf w}}_k}$, respectively, ${{{\bf{\mathord{\buildrel{\lower3pt\hbox{$\scriptscriptstyle\smile$}} \over v} }}}^{\left( i \right)}}$ is the true spectral envelope of instrument $i$, $\alpha$ denotes the mixing weight to control the strength of the constraint imposition, $f$ and $t$ denote the index of frequency bin and time frame, respectively, and $l$ denotes the iteration index. Note that Eq. \ref{Eq11} and \ref{Eq12} are the normalization stages for the spectral bases and Eq. \ref{Eq13} is based on the generalized Dirichlet prior \cite{park2017exploiting}. 

\subsubsection{Spectral envelope of bases}
The spectral envelope of ${\hat {\bf{w}}_k}$ can be obtained as
\begin{equation}	
\label{Eq14}
{{\bf{r}}^{\left( k \right)}} \leftarrow IFFT\left[ {{{\left\{ {{{\hat {\bf{w}}}_k}} \right\}}^2}} \right]
\end{equation}
\begin{equation}	
\label{Eq15}
\begin{aligned}
{{\bf{a}}^{\left( k \right)}} &= {\left( {a_0^{\left( k \right)},\,a_1^{\left( k \right)},\,...\,,\,a_M^{\left( k \right)}} \right)^T} \\
 &\leftarrow LevinsonDurbin\left( {{{\bf{r}}^{\left( k \right)}}} \right)
\end{aligned}
\end{equation}
\begin{equation}	
\label{Eq16}
v_f^{\left( k \right)} \leftarrow \left| {\frac{{{{\eta ^{\left( k \right)}}}}}{{1 - \sum\limits_{m = 1}^M {\left\{ {{a_m}\exp \left( { - {\bf{i}}2\pi \frac{f}{F}} \right)} \right\}} }}} \right|
\end{equation}
\begin{equation}	
\label{Eq17}
e_f^{\left( k \right)} \leftarrow \frac{{\hat {\bf{W}}_{f,k}}}{{v_f^{\left( k \right)}}}
\end{equation}
where ${{\bf{r}}^{\left( k \right)}}$ contains the autocorrelations of $IFFT\left[ {{{\hat {\bf{w}}}_k}} \right]$, ${{\eta ^{\left( k \right)}}}$ is the normalization constant to make ${|| {{{\bf{v}}^{\left( k \right)}}} ||_1} = 1$, and $LevinsonDurbin\left( { \bullet } \right)$ is the function that calculates the $M+1$ dimensional vector ${{\bf{a}}^{\left( k \right)}}$ of LPC coefficients by means of Levinson-Durbin recursion. Note that the imaginary number ${\bf{i}}$ is differentiated from the instrument index $i$, and ${a_0^{\left( k \right)}}=1$ by the definition of LPC.

\subsubsection{True spectral envelope of an instrument}
The true spectral envelope ${{{\bf{\mathord{\buildrel{\lower3pt\hbox{$\scriptscriptstyle\smile$}} 
\over v} }}}^{\left( i \right)}}$ is computed through the similar step. We assume that an audio segment ${{\mathord{\buildrel{\lower3pt\hbox{$\scriptscriptstyle\smile$}} 
\over x} }_i}$ is given for all $i$. First, it is converted to a magnitude spectrogram ${\bf{\mathord{\buildrel{\lower3pt\hbox{$\scriptscriptstyle\smile$}} 
\over X} }}$ and then the spectral envelopes ${{{\bf{\mathord{\buildrel{\lower3pt\hbox{$\scriptscriptstyle\smile$}} 
\over v} }}}_t}$ of each frame ${{{\bf{\mathord{\buildrel{\lower3pt\hbox{$\scriptscriptstyle\smile$}} 
\over x} }}}_t}$ is calculated. Then they are averaged as
\begin{equation}	
\label{Eq18}
{{{\bf{\mathord{\buildrel{\lower3pt\hbox{$\scriptscriptstyle\smile$}} \over v} }}}^{\left( i \right)}} = \frac{{\sum\limits_t {{{ \left\Vert {{{{\bf{\mathord{\buildrel{\lower3pt\hbox{$\scriptscriptstyle\smile$}} \over x} }}}_t}} \right\Vert }_1}{{{\bf{\mathord{\buildrel{\lower3pt\hbox{$\scriptscriptstyle\smile$}} \over v} }}}_t}} }}{{{{ \left\Vert {\sum\limits_t {{{ \left\Vert {{{{\bf{\mathord{\buildrel{\lower3pt\hbox{$\scriptscriptstyle\smile$}} \over x} }}}_t}} \right\Vert }_1}{{{\bf{\mathord{\buildrel{\lower3pt\hbox{$\scriptscriptstyle\smile$}} \over v} }}}_t}} } \right\Vert }_1}}}.
\end{equation}

\subsubsection{Signal reconstruction} 
After the iteration, the estimated spectrograms of each instrument are reconstructed as
\begin{equation}	
\label{Eq19}
{\hat {\bf{X}}_i} = \sum\limits_{k \in {\Phi _i}} {{{\bf{w}}_k}{{\bf{h}}_k}}
\end{equation}
where ${\hat {\bf{X}}_i}$ denotes the estimated spectrogram of instrument $i$. These spectrograms are converted to time domain signals by means of inverse short-time Fourier transform.


\section{Proposed blind approach}
\begin{figure}[t]	
  \centering
  \centerline{\includegraphics[width=3.25in]{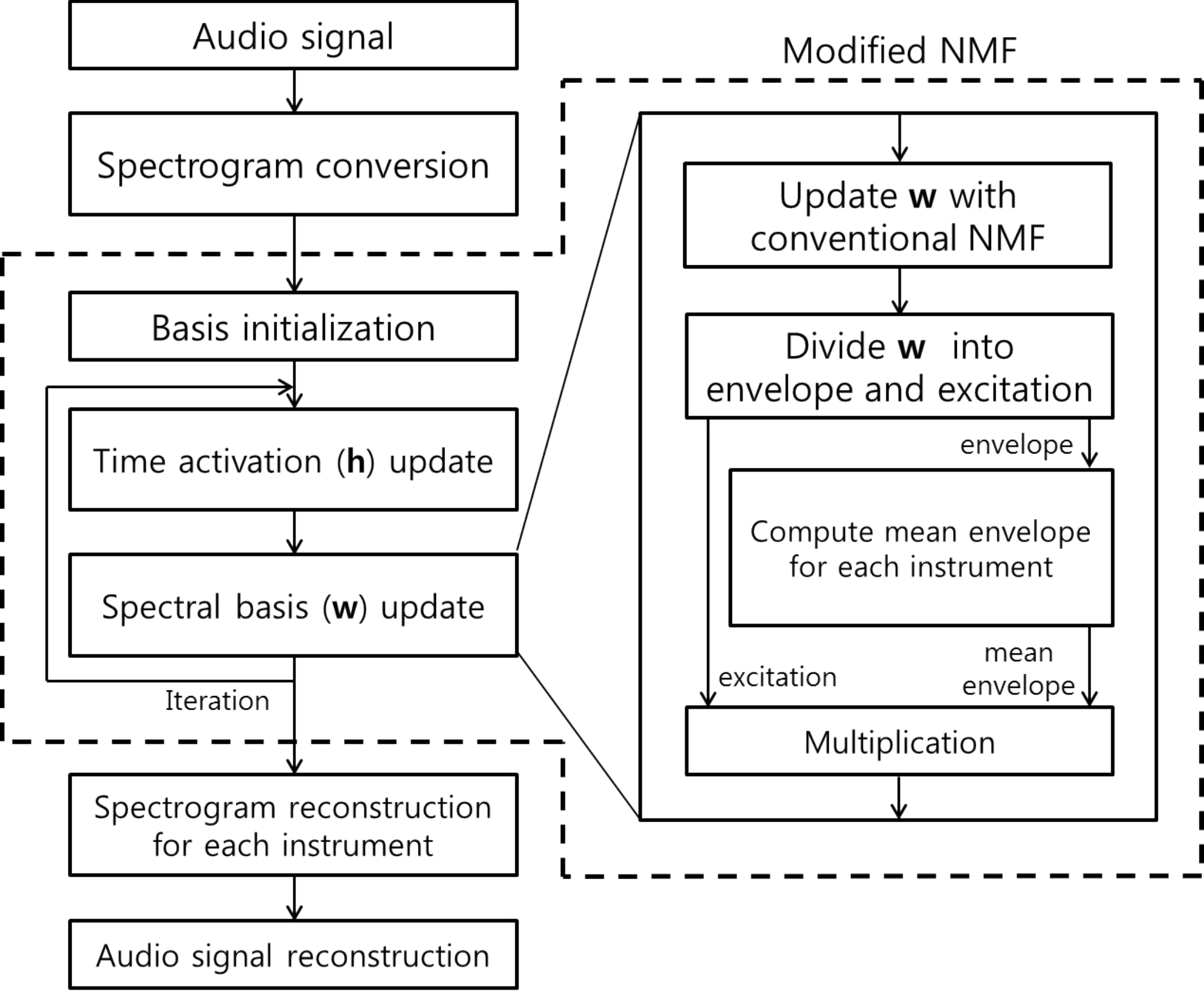}}
  \caption{Overview of the proposed blind approach.}
  \label{Fig3_overview}
\end{figure}

In this section, we describe how we extend the informed approach to the blind scenario. As similar to the previous approach, it is mainly based on the NMF with the generalized Dirichlet prior. Figure \ref{Fig3_overview} shows the structural overview of the proposed method. Similar to the informed approach, the input signal is converted into a spectrogram and decomposed using the NMF to minimize the KL divergence between $\bf{X}$ and $\tilde{\bf{X}}$. During the iteration, the spectral envelopes of the bases belong to the same group are averaged. Then the mean envelope is applied to the bases in the next step. This process can be represented in the mathematical formula as
\begin{equation}	
\label{Eq20}
{\hat {\bf{H}}_{k,t}} \leftarrow \frac{{{\bf{H}}_{k,t}^{\left( l \right)}\sum\limits_f {\left\{ {{\bf{W}}_{f,k}^{\left( l \right)}{{{{\bf{X}}_{f,t}}} \mathord{\left/
 {\vphantom {{{{\bf{X}}_{f,t}}} {{{\tilde {\bf{X}}}_{f,t}}}}} \right.
 \kern-\nulldelimiterspace} {{{\tilde {\bf{X}}}_{f,t}}}}} \right\}} }}{{\sum\limits_{f'} {{\bf{W}}_{f',k}^{\left( l \right)}} }}
\end{equation}
\begin{equation}	
\label{Eq21}
{\tilde {\bf{W}}_{f,k}} \leftarrow \frac{{{\bf{W}}_{f,k}^{\left( l \right)}\sum\limits_t {\left\{ {{\bf{H}}_{k,t}^{\left( {l + 1} \right)}{{{{\bf{X}}_{f,t}}} \mathord{\left/
 {\vphantom {{{{\bf{X}}_{f,t}}} {{{\tilde {\bf{X}}}_{f,t}}}}} \right.
 \kern-\nulldelimiterspace} {{{\tilde {\bf{X}}}_{f,t}}}}} \right\}} }}{{\sum\limits_{t'} {{\bf{H}}_{k,t'}^{\left( {l + 1} \right)}} }}
\end{equation}
\begin{equation}	
\label{Eq22}
{\bf{H}}_{k,t}^{\left( {l + 1} \right)} \leftarrow {\hat {\bf{H}}_{k,t}}\sum\limits_f {{{\tilde {\bf{W}}}_{f,k}}}
\end{equation}
\begin{equation}	
\label{Eq23}
{\hat {\bf{W}}_{f,k}} \leftarrow \frac{{{{\tilde {\bf{W}}}_{f,k}}}}{{\sum\limits_f {{{\tilde {\bf{W}}}_{f,k}}} }}
\end{equation}
\begin{equation}	
\label{Eq24}
{\overline {\bf{v}} ^{{\Phi _i}}} \leftarrow \frac{{\sum\limits_{k \in {\Phi _i}} {{\nu _k}{{\bf{v}}^{\left( k \right)}}} }}{{\sum\limits_{k \in {\Phi _i}} {\sum\limits_{m = 1}^M {{\nu _k}v_m^{\left( k \right)}} } }}
\end{equation}
\begin{equation}	
\label{Eq25}
{\bf{w}}_k^{\left( {l + 1} \right)} \leftarrow  \beta {{\hat {\bf w}}_k} + \left( 1-\beta \right) {\overline {\bf{v}} ^{{\Phi _i}}} \odot {{\bf{e}}^{\left( k \right)}}
\end{equation}
where ${\overline {\bf{v}} ^{{\Phi _i}}}$ is the average spectral envelope for instrument $i$, ${{\nu _k}}$ is the weight of ${{{\bf{v}}^{\left( k\right)}}}$ for the weighted mean, and $\beta$ denotes the mixing weight to control the strength of the constraint imposition. ${{\nu _k}}$ is to be heuristically determined via optimization process presented in the next section.

Note that the Eq. \ref{Eq20}, \ref{Eq21}, \ref{Eq22}, \ref{Eq23} are identical to Eq. \ref{Eq9}, \ref{Eq10}, \ref{Eq11}, \ref{Eq12}, respectively. The true spectral envelope extracted from the given instrument sound is replaced by the group-wise average envelope ${\overline {\bf{v}} ^{{\Phi _i}}}$. Because the proposed blind source separation method is an unsupervised separation method, the ground truth spectral envelopes are not assumed to be given in advance. We have made the bases in a group have a unified envelope, and the envelope converges to the ground truth envelope. Technically, there has been no technique that only estimates the envelopes of the mixed instruments, hence, we have utilized the NMF algorithm that simultaneously learns the envelope and the excitation. Finally, the spectrograms of each instrument are reconstructed in the same way we did in the previous informed approach, followed by the inverse short-time Fourier transform. 

\subsection{Sparse initialization}
\begin{figure}[t]	
  \centering
\begin{subfigure}[]{0.48\textwidth}
\centering
\includegraphics[width=2.6in]{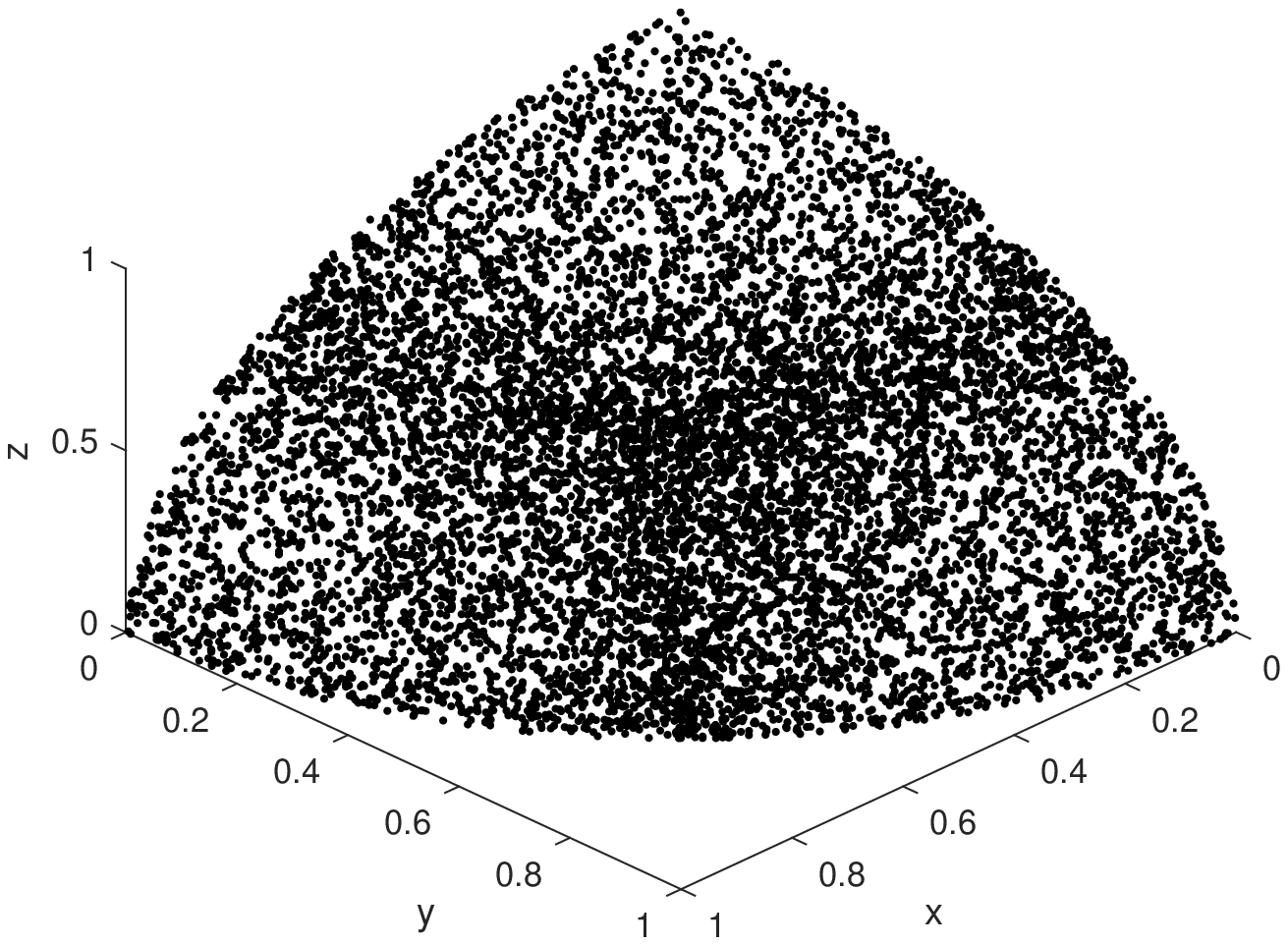}
\subcaption{}
\end{subfigure}
\begin{subfigure}[]{0.48\textwidth}
\centering
\includegraphics[width=2.6in]{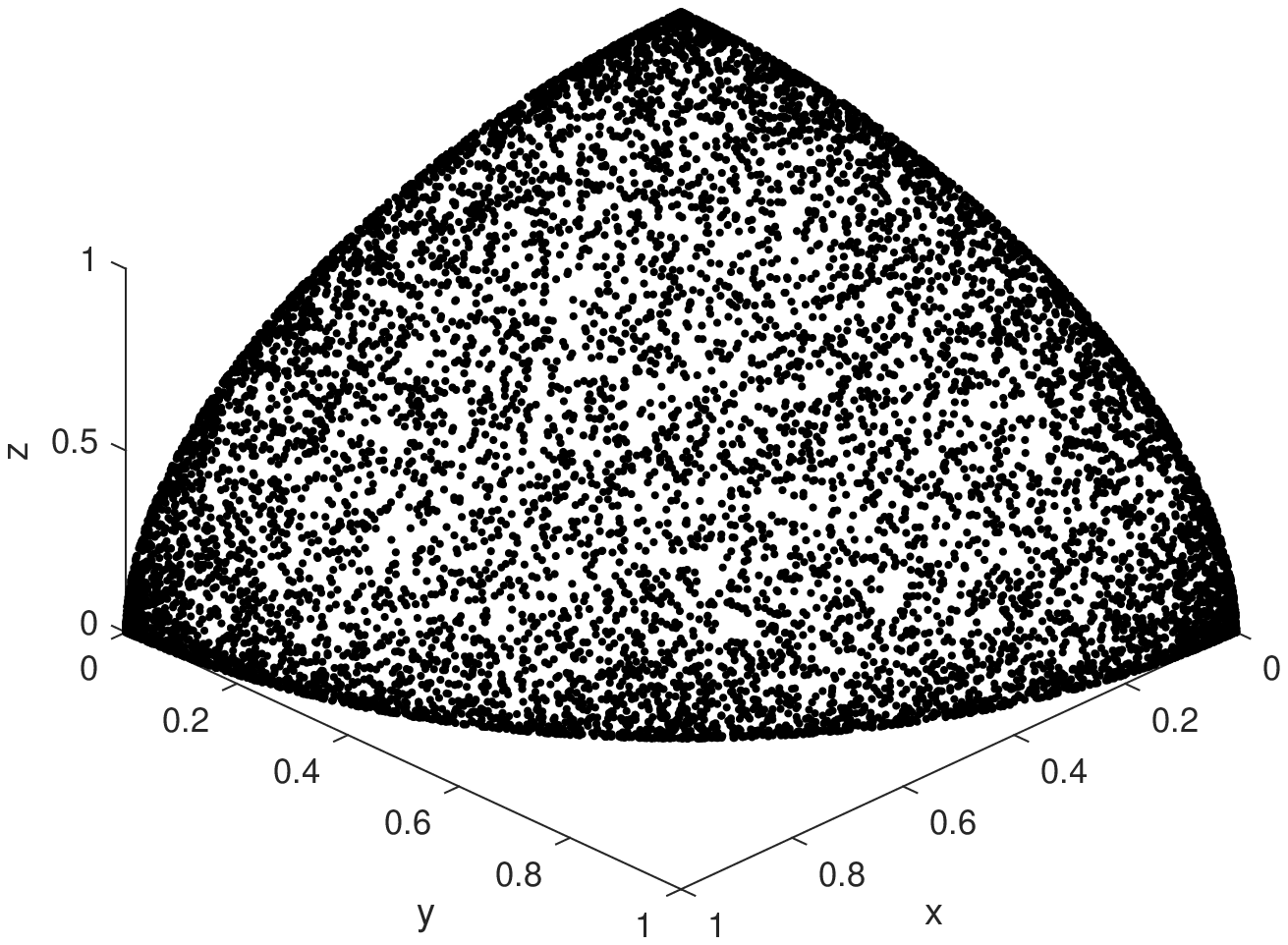}
\subcaption{}
\end{subfigure}
  \caption{Distribution of directions of randomly generated three-dimensional vectors. The vectors are $l_2$ normalized for the visualization.}
  \label{Fig4_direction}
\end{figure}

It has been observed in our previous research \cite{park2017exploiting} that the initialization of the spectral bases can result in the difference of the separation performance. Most of the matrix decomposition-based source separation methods use a Gaussian random noise \cite{spiertz2009source} or uniform random sequence \cite{park2015harmonic}. When we visualize the distribution of their directions as in Figure \ref{Fig4_direction} (a), it can be confirmed that few of them have sparse characteristics. However, the spectral bases often show sparse characteristic, and it can be helpful to initialize them with sparse vectors for the fast convergence as well as performance improvement. Our strategy to generate sparse vectors is to perform elementwise square operation with the uniform random sequence. It makes the sparsity of the bases amplified as we can see in Figure \ref{Fig4_direction} (b). We only utilize the square operation; however, the effect of higher order power operation needs to be further investigated in the future. Using this novel initialization method, we can obtain additional performance gain.

\section{Performance evaluation}
In this section, the performance of the proposed method is evaluated and compared with the baseline algorithms of FASST of Ozerov {\it et al.} and Spiertz and Gnann's methods. For the objective comparison of the performances, we use an audio dataset of real recordings and the representative performance indicators. Before we compare the performances of the methods, we first optimize ${{\nu _k}}$ using an external dataset.

\subsection{RWC dataset}
For the performance evaluation, we have used real-world computing (RWC) music database (musical instrument sound database) \cite{goto2003rwc}. It consists of audio clips of 50 musical instruments, each of which contains variations of playing styles, pitch, and dynamics. The name and the playing styles of the instruments we have used for the experiment are listed in Table \ref{Table1_RWC}. The last two columns represent the hi-hat and snare drum sounds. Note that the RWC database provides three variations in the instrument manufacturer and the performer, and that we have used different variation from the test data with the same playing style for the spectral envelope training. As the second variation of recorder contains the sounds of alto recorder whereas the first variation contains the sounds of soprano recorder, it can occur mismatch in the spectral envelope. Similarly, the first and the second variations of harmonica contain blues harp harmonica and chromatic harmonica, respectively. However, because they show similarity in timbre, we have used the second variations for the envelope training. 

\begin{table*}[] 
\renewcommand{\arraystretch}{1.2}
\caption{Experimental parameters}
\label{Table1_RWC}
\centering
\begin{tabular}{|K{0.7in}|K{2.0in}|K{0.99in}|K{1.1in}|K{0.6in}|}
\hline
Instrument No.	&Instrument name	&Variation No. (Test)	&Variation No. (Training)	&Style\\
\hline
1	&Pianoforte	&1	&2	&NOF\\
\hline
2	&Electric Piano (Hard Tone)	&1	&2	&NOF\\
\hline
7	&Accordion (Keyboard)	&1	&2	&NOF\\
\hline
8	&Harmonica	&1	&2	&NOF\\
\hline
9	&Classic Guitar (Nylon String)	&1	&2	&AFF\\
\hline
13	&Electric Guitar	&1	&2	&LFF\\
\hline
14	&Electric Bass	&1	&2	&PNF\\
\hline
18	&Contrabass (Wood Bass)	&1	&2	&NOF\\
\hline
21	&Trumpet	&1	&2	&M9F\\
\hline
22	&Trombone	&1	&2	&C1F\\
\hline
23	&Tuba	&1	&2	&NOF\\
\hline
24	&Horn	&1	&2	&GNM\\
\hline
27	&Tenor Sax	&1	&2	&NOF\\
\hline
29	&Oboe	&1	&2	&NOF\\
\hline
30	&Bassoon (Fagotto)	&1	&2	&NOF\\
\hline
31	&Clarinet	&1	&2	&NOF\\
\hline
32	&Piccolo	&1	&2	&NOF\\
\hline
33	&Flute	&1	&2	&NOF\\
\hline
34	&Recorder	&1	&2	&NOF\\
\hline
41	&Concert Drums 2	&2	&3	&Y3NO3\\
\hline
41	&Concert Drums 2	&2	&3	&SD1F3\\
\hline
\end{tabular}
\end{table*}

To generate mixtures, we have cut an audio clip to multiple segments, each of which contains the sound of a single note. Then we have mixed the sounds of two instruments to have 10-second duration with 20 notes (10 notes per an instrument) placed at random positions. As drum sounds do not contain pitch information, a sound excerpt is selected and repeated ten times when they are mixed.

\subsection{Evaluation metrics}
We have used the signal-to-distortion ratio (SDR), signal-to-interference ratio (SIR), and signal-to-artifact ratio (SAR) to measure the quantitative performance of the separation methods. They are calculated using the BSS\_EVAL toolbox (http://bass-db.gforce.inria.fr /bss\_eval/) supported by \cite{vincent2006performance}. They are mathematically defined as
\begin{equation} 
\label{Eq26}
SDR = 20{\log _{10}}\left( {\frac{{|| {{s_{target}}} ||}}{{|| {{s_{interf}} + {s_{artif}}} ||}}} \right)
\end{equation}
\begin{equation}
\label{Eq27}
SIR = 20{\log _{10}}\left( {\frac{{|| {{s_{target}}} ||}}{{|| {{s_{interf}}} ||}}} \right)
\end{equation}
\begin{equation}
\label{Eq28}
SAR = 20{\log _{10}}\left( {\frac{{|| {{s_{target}} + {s_{interf}}} ||}}{{|| {{s_{artif}}} ||}}} \right)
\end{equation}
where ${s_{target}}$, ${s_{interf}}$, and ${s_{artif}}$ denote the target sound, interference, and artifact, respectively. As SIR and SAR have a trade-off relationship in the performance, we consider SDR as the representative performance value.

\subsection{Experimental settings}
FASST 2.0 provides many options to perform the source separation. Among the options, we assumed instantaneous mixing scenario with the fixed adaptability, while the rest options are set adaptive. Two types of time-frequency representations, {\it erb} (equivalent rectangular bandwidth) and {\it stft} (short-time Fourier transform), are tested as its options. Also, we evaluate Spiertz and Gnnan's mel-frequency cepstrum coefficient (MFCC) clustering-based method, NMF clustering-based method, and hierarchical clustering-based method.

Table \ref{Table2_parameters} shows the evaluation parameters used for the experiment of the proposed method. The optimal number of bases may be dependent on the number of notes of an instrument. In our experiment, the number of notes in a mixture is 20 in total; however, it is impossible to recognize it in advance. In addition, a single note often requires multiple bases to be thoroughly reconstructed. For this reason, we use 40 bases in the first experiment, and later we observe and compare the performance transitions when the different number of bases is used. The LPC order is empirically determined to 4, but once it was above 2 and below 10, the separation quality did not vary dramatically. Since both our methods and the Spiertz and Gnann's methods require the iteration, the number of iterations is fixed to 100 for all methods for the fair comparison. The mixing weight $\alpha$ linearly increases from 0 to 1 with a step size of 0.01, whereas $\beta$ is fixed to zero in all cases.

\begin{table}[!t] 
\renewcommand{\arraystretch}{1.3}
\caption{Experimental parameters}
\label{Table2_parameters}
\centering
\begin{tabular}{|K{1.8in}|K{1.2in}|}
\hline
Parameter & Value \\
\hhline{|=|=|}
Sampling rate (Hz) & 44,100 \\
\hline
Frame size / Hop size & 4096 / 1024\\
\hline
Number of iterations & 100 \\
\hline
Number of bases (per instrument) & 20, 40, 100 \\
\hline
LPC order ($M$) & 4 \\
\hline
\end{tabular}
\end{table}

\subsection{Weight optimization}
In this subsection, the weight ${{\nu _k}}$ in Eq. \ref{Eq24} is optimized using the ten pieces in the Bach 10 dataset \cite{duan2011soundprism}. The dataset contains real recordings of four instruments; violin, clarinet, saxophone, and bassoon. We have generated a total of sixty cases (six for each piece) where two out of four instrument sounds are linearly mixed. Note that the sounds are amplified or suppressed to have same energies in the mixture. We have analyzed the case where the weight is a function of mean activation ${{{\left\Vert {{{\bf{h}}_k}} \right\Vert}_1}}$. The assumption is that the bases of which spectral envelopes are close to the actual ones will have larger time activity in average.

\begin{figure}[t]	
  \centering
  \centerline{\includegraphics[width=3.3in]{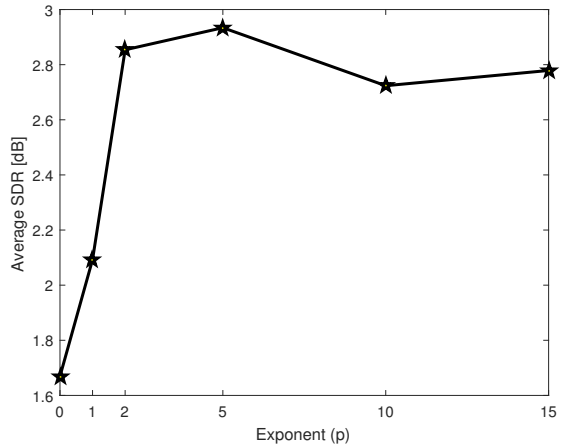}}
  \caption{Average SDR value with the increase of exponent $p$.}
  \label{Fig5_exponent}
\end{figure}

Figure \ref{Fig5_exponent} shows the performance transitions with the increase of the exponent $p$. Here, it is assumed that ${{\nu _k}} = {\left\{ {{{\left\Vert {{{\bf{h}}_k}} \right\Vert}_1}} \right\}^p}$. The envelopes are averaged with equal weights when $p=0$. It can be observed that the average SDR does not fluctuate severely when $p>2$ and is maximized when $p=5$. We use this value as the weight in the rest of the experiments.

\subsection{Performance comparison}
Table \ref{Table3_performance} shows the performance of the conventional methods and the proposed methods. In this experiment, a total of 80 bases (40 for each instrument) are used for our methods and the Spiertz and Gnann's methods. The baseline algorithms of the FASST show the lowest performance in both SIR and SAR, resulting in the lowest SDR. On the contrary, the proposed informed approach shows the highest SIR, SAR, and SDR. This may be due to the aid of side-information which other methods do not use. Except for the informed approach, the proposed blind approach shows the highest SDR performance by showing the highest SIR preserving high SAR. Spiertz and Gnann's method also shows high SIR and SAR values, but their overall performance is severely degraded. For the further investigations on the performance comparison, we perform additional experiments by which we can examine the effect of the number of bases and the sparse initialization. 

\begin{table}[!t] 
\renewcommand{\arraystretch}{1.3}
\caption{Performances measured with the RWC database (dB)}
\label{Table3_performance}
\centering
\begin{tabular}{|K{1.2in}||K{0.5in}|K{0.5in}|K{0.5in}|}
\hline
 & SDR & SIR & SAR \\
\hhline{|=#=|=|=|}
FASST (stft) &			-0.22&	3.14&	7.65 \\
\hline
FASST (erb) &			-0.28&	2.93&	7.80\\
\hline
Spiertz (hierarchical)&	2.71&	6.44&	10.15\\
\hline
Spiertz (MFCC) &		2.72&	6.91&	9.89\\
\hline
Spiertz (NMF) &			2.88&	6.51&	10.18\\
\hline
Proposed (Informed)&	5.55&	9.84&	11.20 \\
\hline
Proposed (Blind)&		3.16&	7.95&	8.88 \\
\hline
\end{tabular}
\end{table}

Table \ref{Table4} shows how the SDRs change according to the variations in the initialization methods and the number of bases. Here, we refer to the initialization method using the uniform random sequence as {\it normal initialization} while the proposed initialization method presented in the section 3.1 is referred to as {\it sparse initialization}. Starting from 20 bases, which is the minimum because it is identical to the number of notes, to 100 bases, which is sufficient, we have tested our algorithms and the Spiertz and Gnann's algorithms. From the results of the proposed informed approach, we can confirm that its performance is not affected by the number of bases. Other blind approaches reveal that their performances tend to increase when we use more bases. In particular, the performance of the proposed blind approach with 20 normal initialized bases is highly enhanced when we increase the number of bases to 40. The sparse initialization partly covers the short of bases by reforming the bases into more meaningful shapes. Regardless of the number of bases, the use of sparse initialization improves the SDR performance. Overall, the proposed blind approaches tend to outperform the Spiertz and Gnann's methods.

\begin{table}[] 
\centering
\renewcommand{\arraystretch}{1.2}
\caption{Effect of the number of bases and the initialization methods}
\label{Table4}
\begin{tabular}{|K{0.8in}|K{0.6in}||K{0.4in}|K{0.4in}|K{0.4in}|}
\hline
\multicolumn{2}{|K{1.4in}||}{\multirow{2}{*}{}} 	& \multicolumn{3}{K{1.5in}|}{Number of bases} \\ \cline{3-5} 
\multicolumn{2}{|K{1.4in}||}{} 				& 20	  &	40	& 100        \\ \hhline{|==#=|=|=|}
\multicolumn{2}{|K{1.4in}||}{Spiertz (hierarchical)} 	& 2.52 &	2.71	& 2.96    \\ \hline
\multicolumn{2}{|K{1.4in}||}{Spiertz (MFCC)} 		& 2.66 &	2.72 &	2.84    \\ \hline
\multicolumn{2}{|K{1.4in}||}{Spiertz (NMF)} 		& 2.46 &	2.88 &	3.03  \\ \hline
\multicolumn{2}{|K{1.4in}||}{Proposed (Informed)}	& 5.56 &	5.55 &	5.59 \\ \hline
\multirow{2}{*}{Proposed (Blind)} & Normal init. 	& 2.50 &	3.05 &	3.14  \\ \cline{2-5} 
                                  			  & Sparse init. 	& 3.02 &	3.16 &	3.34  \\ \hline
\end{tabular}
\end{table}

\section{Conclusion}
In this paper, we presented novel source separation methods that exploit spectral envelope information in the spectrogram decomposition framework. To this end, the proposed method divided basis vectors into the spectral envelopes and the corresponding excitations using linear prediction. By enforcing constraints to the envelope part of the bases in every iteration, we could shape the bases to represent the desired instruments. Our first research assumed that we could obtain the instrument-specific envelopes through the provided audio clips. In the second part of our study, we extended the informed approach to the situations where the acquisition of the true instrument envelopes is impossible. We evaluated the performance of the proposed methods and verified that our methods outperform other conventional methods.

\vspace{6pt} 

\bibliography{references}
\bibliographystyle{plain}

\end{document}